\documentclass[%
aip,
rsi,
twocolumn,
amsmath,amssymb,
groupedaddress,
reprint,
]{revtex4-1}
\usepackage{amsmath,amssymb,amsfonts,bm}
\usepackage{graphicx,epstopdf}
\usepackage{color}
\usepackage{extarrows}
\usepackage{enumitem}
\usepackage{empheq}
\usepackage{chngcntr}
\usepackage{appendix}
\usepackage{diagbox}

\renewcommand{\d}{{\rm d}}

\newcommand{\w}{\omega}
\newcommand{\wti}{\widetilde}
\newcommand{\ti}{\tilde}
\newcommand{\B}{\mbox{\tiny B}}

\newcommand{\s}{\mbox{\tiny S}}

\newcommand{\tS}{\mbox{\tiny S}}

\newcommand{\T}{\mbox{\tiny T}}

\newcommand{\eff}{\rm eff}

\newcommand{\la}{\langle}
\newcommand{\ra}{\rangle}

\newcommand{\as}{\alpha}
\newcommand{\asb}{\alpha}
\newcommand{\h}{\hat}
\newcommand{\ini}{t}

\newcommand{\cc}{\bm{\chi}}

\newcommand{\iit}{\int_{0}^{t}\!{\rm d}\tau\,}
\newcommand{\iitp}{\int_{0}^{t}\!{\rm d}\tau'\,}

\newcommand{\bo}{\bm{\Omega}}
\newcommand{\bu}{\bm{u}}

\newcommand{\bW}{\bm{W}}
\newcommand{\bv}{\bm{V}}

\newcommand{\bp}{{\bm{\phi}}_{\alpha}}

\newcommand{\wbpi}{\widetilde{\bm{\phi}}}
\newcommand{\bC}{\bm{C}}
\newcommand{\bt}{\textbf{T}}

\newcommand{\App}[1]{Appendix}
\newcommand{\Sec}[1]{Sec.\,\ref{#1}}
\newcommand{\nl}{\nonumber \\}
\newcommand{\be}{\begin{equation}}
\newcommand{\ee}{\end{equation}}
\newcommand{\bsube}{\begin{subequations}}
\newcommand{\esube}{\end{subequations}}
\newcommand{\eq}[1]{Eq.\,(\ref{#1})}
\newcommand{\Eq}[1]{Eq.\,(\ref{#1})}
\newcommand{\eqs}[1]{Eqs.\,(\ref{#1})}
\newcommand{\Fig}[1]{Fig.\,\ref{#1}}

\newcommand{\obarplus}{\mbox{\tiny$\oplus$}}
\newcommand{\obarminus}{\mbox{\tiny$\ominus$}}

\newcommand{\RN}[1]{%
  \textup{\uppercase\expandafter{\romannumeral#1}}%
}

\begin{document}
\title{Multimode Brownian oscillators: Exact solutions to heat transport}
\author{Xin-Hai Tong}
\affiliation{CAS Key Laboratory of Precision and Intelligent Chemistry,
University of Science and Technology of China, Hefei, Anhui 230026, China}
\affiliation{School of Physics and Technology, Wuhan University, Wuhan, Hubei 430072, China}
\author{Hong Gong}
\affiliation{CAS Key Laboratory of Precision and Intelligent Chemistry ,
University of Science and Technology of China, Hefei, Anhui 230026, China}
\author{Yao Wang}
\email{wy2010@ustc.edu.cn}
\affiliation{Hefei National Research Center for Physical Sciences at the Microscale,
University of Science and Technology of China, Hefei, Anhui 230026, China}
\author{Rui-Xue Xu}
\email{rxxu@ustc.edu.cn}
\affiliation{Hefei National Research Center for Physical Sciences at the Microscale,
University of Science and Technology of China, Hefei, Anhui 230026, China}
\author{YiJing Yan}
\affiliation{Hefei National Research Center for Physical Sciences at the Microscale,
University of Science and Technology of China, Hefei, Anhui 230026, China}

\date{\today}

\begin{abstract}
In this work, we investigate the multimode Brownian oscillators in nonequilibrium scenarios with multiple reservoirs at different temperatures.
For this purpose, an algebraic method is proposed.
This approach gives the exact time--local equation of motion for reduced density operator, from which
we can easily extract not only the reduced system but also hybrid bath dynamical information.
The resulted steady--state heat current is found numerically consistent with another discrete imaginary--frequency method followed by the Meir--Wingreen's formula.
It is anticipated that the development in this work would constitute an indispensable component to nonequilibrium statistical mechanics for open quantum systems.
\end{abstract}
\maketitle

\section{Introduction}
Open quantum systems play pivotal roles in diversed fields such as
nuclear magnetic resonance,\cite{Rei82,Sli90,Van051037}
condensed matter and material physics,\cite{Bor85,Kli97,Ram98}
high energy physics,\cite{Aka15056002,Bla181,Miu20034011}
quantum optics,\cite{Hak70,Lou73,Scu97}
chemical and biological physics, \cite{Nit06,Cre13253601,Fan22084119}
and nonlinear spectroscopies. \cite{She84,Yan885160,Muk95}
In most of these studies, the system and its environment constitute a thermodynamic composite.
Thermal effects dictate the
system--environment entanglement, which is intimately related to the thermodynamic and
transport properties.\cite{Kat16224105,Son17064308,Gon20154111,Koy22014104,Wan22044102}
Practically, these properties are
under focus in manipulating mesoscopic nanodevices.\cite{Hsi22125018,Ang974041,Gar04,Cle101155,Soa14825,Har201184}

For example, the real-time evolution of heavy quarkonium in the quark-gluon plasma was investigated
by means of quantum Brownian motion theory
to highlight dynamical mechanism of the relative motion of the quarkonium state.\cite{Miu20034011}
Another example is that when considering a point charge emitting radiation in an electromagnetic field,
the internal degree of freedom of a moving atom can be modeled by a harmonic oscillator coupled to a scalar field.
Following the same coupling way as scalar electrodynamics,
the pathologies of radiation reactions can be recast as non-Markovian dynamics of a Brownian oscillator (BO)
in an environment.\cite{Hsi22125018}   
In terms of quantum decoherence and measurement,
BO was also introduced as one of toy models to study universalities in the phenomenology of decoherence.\cite{Ang974041}
These continuous studies on BO indicate the significance of this model can be never underestimated.

 As a typical open quantum system, BO is the simplest and exactly solvable model.
 It serves as an elementary consideration in various studies.\cite{Cal83587,Gra88115,Yan05187,Wei21,Pur16062114,Hof17123037,Far19012107}
Exact real-time dynamics of BO has been widely investigated.\cite{Haa852462,Unr891071,Hu922843,Hal962012,Kar97153,For01105020,Xu09074107}
 Concerning thermodynamics,
 properties of the reduced BO system had been shown that
the steady state could be expressed as the Gibbs state with a renormalized Hamiltonian.\cite{Hua22023141}
On the other hand, it should be noticed that
besides the reduced system, the hybrid bath properties need also be taken into account
for the underlying entangled effects.\cite{Gon20214115,Wan22044102,Gon22054109}

In this work, we develop an algebraic approach to obtain the  equation of motion (EOM) for the multimode BO system, from which we can easily extract both the reduced system and hybrid bath dynamical information.
Furthermore, we exploit the well-established EOM to study the heat transport problem.
We also propose another discrete imaginary--frequency (DIF) method
 followed by the Meir--Wingreen's formula.
Both the algebraic and DIF methods are consistent with each other in evaluating the steady--state heat current.

The remainder of this paper is organized as follows.
 In \Sec{thsec2} we present the multimode BO system model and the exact EOM for dynamics.
 Especially, we emphasize how to extract the hybrid bath information and to evaluate the heat current.
More theoretical details are given in Appendix.
Numerical results are demonstrated in \Sec{num}.
We summarize the paper in \Sec{thsec4}.
 Throughout this paper we set $ \hbar=1 $ and $ \beta_\alpha=1/(k_{B}T_{\alpha}) $ with  $ k_{B} $ being the Boltzmann constant and $ T_{\alpha} $ the temperature of the $\alpha$-reservoir.

\section{Multimode Brownian Oscillator}\label{thsec2}
\subsection{Hamiltonian and EOM}
Let us start with the total system--bath Hamiltonian, $ H_{\T}=H_{\s}+H_{\s \B}+h_{\B} $.
The bath  $ h_{\B}=\sum_{\as}h_{\as} $ is assumed to be non--interacting with different temperatures $ \beta_{\as} $.
The BO model adopts
 \begin{align}\label{Hspq}
 H_{\s}=\sum_{u}\frac{\hat p_{u}^{2}}{2m_{u}}+\frac{1}{2}\sum_{uv}k_{uv}\hat q_u\hat q_v,
 \end{align}
 and
\begin{align}\label{2}
H_{\s\B}=\sum_{\alpha u j}\kappa_{\as uj}\hat q_u\hat{x}_{\as j}
\equiv\sum_{\as u}\hat Q_u\hat F_{\alpha u},
\end{align}
with
\begin{align}
&\hat{Q}_{u}=(m_{u}k_{uu})^{\frac{1}{4}}\h{q}_{u}.
\end{align}
This is the dimensionless disspative system mode that coupled to
the random force $ \hat{F}_{\as u} $ as specified in \eq{2}.
Note also $ \hat{P}_{u}=(m_{u}k_{uu})^{-\frac{1}{4}}\h{p}_{u} $ so that $ [\hat{Q}_{u},\hat{P}_{v}]=i\delta_{uv} $.
Thus \Eq{Hspq} is recast as
\begin{align}
H_{\s}=\frac{1}{2}\sum_{u}\Omega_{u}\hat{P}^{2}_{u}+\frac{1}{2}\sum_{uv}V_{uv}\hat{Q}_{u}\hat{Q}_{v}
\end{align}
with
	\begin{align}
	\Omega_{u}\equiv\sqrt{\frac{k_{uu}}{m_{u}}}\quad
{\rm and}\quad
	V_{uv}\equiv\frac{k_{uv}}{\sqrt[4]{m_{u}m_{v}k_{uu}k_{vv}}}.
	\end{align}

In order to investigate some system--bath entangled properties such as the heat current, we introduce an auxiliary quantity
\begin{align}\label{defrho}
\ti{\rho}_{\alpha u}(t) \equiv \text{tr}_{\B}[\hat F_{\alpha u}\rho_{\T}(t)].
\end{align}
By applying the Liouville equation to the total density operator, $ \dot{\rho}_{\T}(t)=-i[H_{\T},\rho_{\T}(t)] $,
we obtain for the reduced $ \rho_{\s}(t)=\text{tr}_{\B}\rho_{\T}(t) $
the EOM as
\begin{align}\label{e67}
\dot{\rho}_{\s}(t)=&-i[H_{\s},\rho_{\s}(t)]-i\sum_{\alpha u}[\hat Q_{u},\ti{\rho}_{\alpha u}(t)].
\end{align}
The above equation exists for any given $ H_{\s} $.
It leads to the exact hierarchical equations of motion for general systems.\cite{Tan20020901}
Focusing on the BO system, \eq{e67} can be evaluated in a closed form with
\begin{align}\label{varrhonew1}
\ti\rho_{\as u}(t)&=\sum_{v}\big[\wti\Gamma_{\alpha uv}(t)\hat{Q}_{v}^{\obarplus} +\Gamma_{\alpha uv}(t) \hat{P}_{v}^{\obarplus}
\nl&\qquad\quad+\ti\zeta_{\alpha uv}(t)\hat{Q}_{v}^{\obarminus}-\zeta_{\alpha uv}(t)\hat{P}_{v}^{\obarminus}\big]\rho_{\s}(t)
\end{align}
where
\begin{align}\label{defop1}
&\hat A^{\obarplus}\hat{O}\equiv\frac{1}{2}\{\hat{A},\hat{O}\}\quad\text{and}\quad \hat A^{\obarminus}\hat{O}\equiv-i[\hat{A},\hat{O}].
\end{align}
Apparently, \Eq{e67} is trace preserving for the reduced density operator,
since ${\rm tr}_{\tS}\dot{\rho}_{\tS}(t)=0$.
The EOM (\ref{e67})--(\ref{defop1}) is  exact and non-Markovian, fully taking account of the  environmental temperature effects.
As an exact approach, it will guarantee the positivity of $\rho_{\tS}$.
Determination of the EOM (\ref{e67})--(\ref{defop1})
is detailed in \App{appb}.

\subsection{Details of the time-dependent coefficients in the EOM}
To specify the involved time-dependent coefficients in \Eq{varrhonew1},
let us first introduce a key quantity
\begin{align}\label{chi}
\bm{\chi}(t)\equiv\cc_{QQ}(t)\equiv\{\chi^{QQ}_{uv}(t)\equiv i\langle  [\hat Q_{u}(t),\hat Q_{v}(0)] \rangle\}
\end{align}
with  $\hat Q_{u}(t)\equiv e^{iH_{\T}t}\hat Q_{u}e^{-iH_{\T}t}$
and the average $ \langle  \cdots \rangle $ over the steady state of total composite
$\rho^{\rm st}_{\T}$ which commutes with $H_{\T}$.
Similarly, we denote
\[
\begin{split}
& \cc_{QP}(t)\equiv\{\chi^{QP}_{uv}(t)\equiv i\langle  [\hat Q_{u}(t),\hat P_{v}(0)]\ra\}=-\dot{\bm\chi}(t)\bo^{-1},
\\
&\cc_{PQ}(t)\equiv\{\chi^{PQ}_{uv}(t)\equiv i\langle  [\hat P_{u}(t),\hat Q_{v}(0)]\ra\}=\bo^{-1}\dot{\bm\chi}(t),
\\
&\cc_{PP}(t)\equiv\{\chi^{PP}_{uv}(t)\equiv i\langle  [\hat P_{u}(t),\hat P_{v}(0)]\ra\}=-\bo^{-1}\ddot{\bm\chi}(t)\bo^{-1},
\end{split}
\]
where $\bo\equiv\{\Omega_{u}\delta_{uv}\}$.
The associated initial values are
$\bm{\chi}(0)=\ddot{\bm{\chi}}(0)=\bf{0}$ and $\dot{\bm{\chi}}(0)=\bm{\Omega}$.

The time evolutions of system coordinates and momentums
can be resolved as\cite{Xu037,Yan05187,Xu09074107}
\begin{align}\label{nonlocal}
\begin{bmatrix} \bm{\hat Q}(t) \\ \bm{\hat P}(t) \end{bmatrix}
= {\bf T}(t)
\begin{bmatrix} \bm{\hat Q}(0) \\ \bm{\hat P}(0) \end{bmatrix}
- \sum_{\alpha}\int_{0}^{t} \d\tau\, {\bf T}(t-\tau)
\begin{bmatrix} \bf{0} \\\bm{\hat{F}}^{\B}_{\alpha}(\tau) \end{bmatrix}.
\end{align}
For compactness, we have introduced vectors, $\bm{\hat{P}}\equiv\{\hat P_u\}$, $\bm{\hat Q}\equiv\{\hat Q_u\}$, $\bm{\hat{F}^{\B}}_{\alpha}\equiv\{\hat{F}^{\B}_{\alpha u}\}$, and
the matrix
\begin{align}\label{tmatirx}
{\bf T}(t)
= \begin{bmatrix}
-\cc_{QP}(t) & \cc_{QQ}(t)
\\      -\cc_{PP}(t) & \cc_{PQ}(t)
\end{bmatrix}= \begin{bmatrix}
\dot{\bm\chi}(t)\bo^{-1} & \bm\chi(t)
\\      \bo^{-1}\ddot{\bm\chi}(t)\bo^{-1} & \bo^{-1}\dot{\bm\chi}(t)
\end{bmatrix}.
\end{align}
By taking the time derivative of \Eq{nonlocal} and eliminating the initial values using itself, the time--local EOM can be obtained as\cite{Xu037,Yan05187,Xu09074107}
\begin{align}\label{local}
\begin{bmatrix} \dot{\bm{\hat Q}}(t) \\ \dot{\bm{\hat P}}(t) \end{bmatrix}= {\bm\Lambda}(t)
\begin{bmatrix} \bm{\hat Q}(t) \\ \bm{\hat P}(t) \end{bmatrix}-\sum_{\alpha}\begin{bmatrix}
\bf{0}\\ \bm{\hat{F}}^{\rm eff}_{\alpha}(t)
\end{bmatrix}.
\end{align}
Here, the matirx $ \bm{\Lambda}(t) $ is resulted as
\begin{align}\label{lammda}
{\bm\Lambda}(t)&= \dot{\bf T}(t){\bf T}^{-1}(t)=\begin{bmatrix}
\     \bf{0}\     &
\     \bo\     \\
\    -\bv-\bm{\wti\Gamma}(t)\      &
\    -\bm{\Gamma}(t)\
\end{bmatrix}
\end{align}
and (denoting $\bm{\wti V}(t)\equiv\bm{\wti\Gamma}(t)+\bv$)
\begin{align}\label{Feff_vc}
\bm{\hat{F}}^{\rm eff}_{\alpha}(t)&=\bm{\hat{F}}^{\B}_{\alpha}(t)+\iit\big[\bm{\wti V}(t)\bm{\chi}(t-\tau)
\nl &\qquad +\bm{\Gamma}(t)\bo^{-1}\dot{\cc}(t-\tau)+\bo^{-1}\ddot{\cc}(t-\tau)\big]\bm{\hat{F}}^{\B}_{\alpha}(\tau),
\end{align}
where 
\begin{subequations}\label{boandbg}
	\begin{align}
	&\bo\bm{\Gamma}\bo^{-1}=\dddot{\cc}(\dot{\cc}\cc^{-1}\dot{\cc}-\ddot{\cc})^{-1}-\ddot{\cc}(\dot{\cc}-\ddot{\cc}\dot{\cc}^{-1}\cc)^{-1},\\
	&\bo\bm{\wti V}=\dddot{\cc}(\cc\dot{\cc}^{-1}\ddot{\cc}-\dot{\cc})^{-1}-\ddot{\cc}(\cc-\dot{\cc}\ddot{\cc}^{-1}\dot{\cc})^{-1}.
	\end{align}
\end{subequations}
They are obtained from \Eq{lammda} by using the time derivative of \Eq{tmatirx} and
\begin{align}
\bt^{-1}(t)
=
\begin{bmatrix}
  \bo(\dot{\cc}-\cc\dot{\cc}^{-1}\ddot{\cc})^{-1}
&  \bo(\ddot{\cc}-\dot{\cc}\cc^{-1}\dot{\cc})^{-1}\bo
\\  (\cc-\dot{\cc}\ddot{\cc}^{-1}\dot{\cc})^{-1}
&    (\dot{\cc}-\ddot{\cc}\dot{\cc}^{-1}\cc)^{-1}\bo
\end{bmatrix}.
\end{align}
Note that all the involved elements in $\bm{\chi}(t)$
and the related $\bm{\Gamma}(t)$ and $\bm{\wti\Gamma}(t)$ are real.

 For the investigation on the heat transport, we shall identify the contribution of each hybridized
$\alpha$--bath to the key functions $\bm{\wti\Gamma}(t)$ and $\bm{\Gamma}(t)$ in terms of
\be\label{separate}
   \bm{\wti\Gamma}(t)=\sum_\alpha \bm{\wti\Gamma}_{\alpha}(t)  \quad {\rm and}\quad
       \bm{\Gamma}(t)=\sum_\alpha     \bm{\Gamma}_{\alpha}(t).
\ee
Here, $\bm{\wti\Gamma}_{\alpha}(t)\equiv\{\wti\Gamma_{\alpha uv}(t)\}$ and
$\bm{\Gamma}_{\alpha}(t)\equiv\{\Gamma_{\alpha uv}(t)\}$ [cf.\,\Eq{varrhonew1}].
They can be attained by %
\begin{align}\label{Gamma_contri}
\begin{bmatrix}
\bm{\wti\Gamma}_{\alpha}(t) &  \bm{\Gamma}_{\alpha}(t)
\end{bmatrix}=-\iit
\begin{bmatrix}
\bp(t-\tau) & {\bf 0}
\end{bmatrix}\bt(\tau)\bt^{-1}(t).
\end{align}
Here $\bp(t)\equiv\{\phi_{\alpha uv}(t)\}$ with
\begin{align}\label{phiauv}
  &\phi_{\alpha uv}(t)\equiv i\langle  [\hat F^{\B}_{\alpha u}(t),\hat F^{\B}_{\alpha v}(0)] \rangle_{\B}
\end{align}
and $\hat F^{\B}_{\alpha u}(t)\equiv e^{ih_{\alpha}t}\hat F_{\alpha u}e^{-ih_{\alpha}t}$.
The average is defined as $ \langle  \cdots \rangle_{\B}\equiv \text{tr}_{\B}[\cdots\,\rho^{\rm cano}_{\B}]$
with $ \rho^{\rm cano}_{\B}=\otimes_{\alpha}\rho^{\B}_{\alpha}(\beta_{\as})$ and
$\rho^{\B}_{\alpha}(\beta_{\as})=e^{-\beta_{\alpha}h_{\alpha}}/\text{tr}_{\B}(e^{-\beta_{\alpha}h_{\alpha}})$
over the canonical ensembles of baths.
We shall also introduce the hybrid bath correlation function
	\begin{align}
	c_{\alpha uv}(t)\equiv\langle \hat{F}^{\B}_{\alpha u}(t)\hat{F}^{\B}_{\alpha v}(0)  \rangle_{\B}.
	\end{align}
With the symmetry property $c^\ast_{\alpha uv}(t)=c_{\alpha vu}(-t)$,
we have $\phi_{\alpha uv}(t)=i[c_{\alpha uv}(t)-c^{\ast}_{\alpha uv}(t)]=-2{\rm Im}c_{\alpha uv}(t)=-\phi_{\alpha vu}(-t)$.
Denote also $r_{\alpha uv}(t)\equiv{\rm Re}[c_{\alpha uv}(t)]=r_{\alpha vu}(-t)$ for later use.
Note that the cross correlations between different baths do not exist.
More details for obtaining \Eq{Gamma_contri} are given in \App{appa}.

Turn now to $\bm{\wti\zeta}_{\alpha}(t)\equiv\{\wti\zeta_{\alpha uv}(t)\}$ and
$\bm{\zeta}_{\alpha}(t)\equiv\{\zeta_{\alpha uv}(t)\}$ in \Eq{varrhonew1}.
Their evaluations can be summarized as
\begin{align}\label{zetafinal}
&\begin{bmatrix}
\bm{\zeta}_{\alpha}(t) &
\bm{\wti\zeta}_{\alpha}(t)
\end{bmatrix}
\nl &
=\begin{bmatrix}
\bm{\wti V}(t) & \bm{\Gamma}(t)\bo^{-1} & \bo^{-1}
\end{bmatrix}
\begin{bmatrix}
\bm{k}^{00}_{\as}(t) & \bm{k}^{01}_{\as}(t)\bo^{-1} \\
\bm{k}^{10}_{\as}(t) & \bm{k}^{11}_{\as}(t)\bo^{-1} \\
\bm{k}^{20}_{\as}(t) & \bm{k}^{21}_{\as}(t)\bo^{-1} \\
\end{bmatrix}
\end{align}
where (with $T$ denoting the transpose)
\be\label{bmkij}
     \bm{k}^{ij}_{\as}(t)=\iit \big[\mathcal{D}_{i}\bu_{\as}(\tau)\mathcal{D}_{j}\cc^{T}(\tau)
       +\mathcal{D}_{i}\cc(\tau)\mathcal{D}_{j}\bu^{T}_{\as}(\tau)\big],
\ee
with
\be\label{bu}
	\bu_{\alpha}(t)\equiv\int_{0}^{t}\!{\rm d}\tau\,\cc(\tau) \bm{r}_{\asb}(t-\tau),
\ee
and the notation $\mathcal{D}_{m}$ for the derivative $d^{m}/d\tau^{m}$.
Detailed derivations are given in \App{appa}.
For the condition of a single oscillator coupled to a single bath,
\Eq{zetafinal} will recover the results in the APPENDIX C of Ref.\,\onlinecite{Xu09074107},
where the expressions had been reorganized via integral by parts.
The asymptotic behaviors had been found there associated with the thermal phase space variances.

\subsection{Entangled system--bath properties and heat transport}
\label{thsec3b}
Consider
the heat transport from the $\alpha$--reservoir
to the local impurity system.  The heat current operator reads
\begin{align}\label{hatI_boson_def}
 \hat J_{\alpha}
&\equiv -\dot h_{\alpha} 
     =-i[H_{\T},h_{\alpha}]
     =-i\sum_{u}\hat Q_{u}[{\hat F}_{\alpha u},h_{\alpha}]
\nl &=-i\sum_{u}\hat Q_{u}[{\hat F}_{\alpha u},H_{\T}]
 =\sum_{u}\hat Q_{u}\dot{\hat F}_{\alpha u}.
\end{align}
Denote $\langle\hat O\rangle_{\ini}={\rm Tr}[\hat O\rho_{\T}(t)]$.
We can evaluate the heat current via
\begin{align}\label{current}
 \langle  \h J_{\alpha}(t) \rangle_{\ini}&=
           -i\sum_{u}\la\hat Q_{u}[{\hat F}_{\alpha u},H_{\T}]\ra_{\ini}
\nl &=     -i\sum_{u}{\rm Tr}\big\{[H_{\T},\rho_{\T}(t)]\h Q_{u}\h F_{\alpha u}\big\}
\nl &\quad -i\sum_{u}{\rm Tr}\big\{\rho_{\T}(t)[H_{\T},\h Q_{u}]\h F_{\alpha u}\big\}
\nl &=\sum_{u}\big[\dot{\wti Q}_{\alpha u}(t)-\Omega_{u}\wti P_{\alpha u}(t)\big],
\end{align}
where [cf.\ \eqs{varrhonew1} and (\ref{B3})]
\bsube
\begin{align}\label{eq27a}
  \wti Q_{\alpha u}&\equiv{\rm tr}_{\s}\big[\h Q_{u}{\ti{\rho}}_{\alpha u}(t)\big]
\nl &=-\zeta_{\alpha uu}(t)+\sum_v\wti\Gamma_{\alpha uv}(t)\big[W^{QQ}_{uv}(t)+Q_u(t)Q_v(t)\big]
\nl &\qquad                    +\sum_v\Gamma_{\alpha uv}(t)\big[W^{QP}_{uv}(t)+Q_u(t)P_v(t)\big],
\end{align}
and
\begin{align}\label{eq27b}
  \wti P_{\alpha u}&\equiv{\rm tr}_{\s}\big[\h P_{u}{\ti{\rho}}_{\alpha u}(t)\big]
\nl &=-\ti\zeta_{\alpha uu}(t)+\sum_v\wti\Gamma_{\alpha uv}(t)\big[W^{PQ}_{uv}(t)+P_u(t)Q_v(t)\big]
\nl &\qquad                       +\sum_v\Gamma_{\alpha uv}(t)\big[W^{PP}_{uv}(t)+P_u(t)P_v(t)\big].
\end{align}
\esube
Involved in Eqs.\,(\ref{eq27a}) and (\ref{eq27b}) are also the first order moments,
$\{Q_u(t)\}$ and $\{P_u(t)\}$, and the second order moments,
$\{W_{uv}^{QQ}\}$, $\{W_{uv}^{QP}\}$, $\{W_{uv}^{PQ}\}$,
and $\{W_{uv}^{PP}\}$.
Their definitions are given in \Eq{1st_def} and \Eq{B3}, respectively.
Thus \Eq{current} can be recast in terms of
\be
  \la\h J_{\alpha}(t) \ra_{\ini}={\rm trace}\big[\dot{\bm J}_{\alpha;Q}(t)-\bo{\bm J}_{\alpha;P}(t)\big]
\ee
with
\bsube
\begin{align}
 \bm J_{\alpha;Q}(t)&=
   \bm{\wti\Gamma}_{\alpha}(t)\big[\bm{W}^{QQ}(t)+\bm{Q}^T(t)\bm{Q}(t)\big]      -\bm{\zeta}_{\alpha}(t)
\nl &\qquad   +\bm{\Gamma}_{\alpha}(t)\big[\bm{W}^{PQ}(t)+\bm{P}^T(t)\bm{Q}(t)\big],\\
 \bm J_{\alpha;P}(t)&=
   \bm{\wti\Gamma}_{\alpha}(t)\big[\bm{W}^{QP}(t)+\bm{Q}^T(t)\bm{P}(t)\big]    -\bm{\wti\zeta}_{\alpha}(t)
\nl &\qquad   +\bm{\Gamma}_{\alpha}(t)\big[\bm{W}^{PP}(t)+\bm{P}^T(t)\bm{P}(t)\big].
\end{align}
\esube
The steady--state heat current is obtained by $t\rightarrow\infty$.

An alternative approach to evaluate the steady--state heat current is via
\be\label{heat_current}
 J_{\alpha}\equiv\la\h J_{\alpha}(t) \ra_{\ini}\big|_{t\rightarrow\infty}=-2\,{\rm Im}\!\int^{\infty}_0\!\!\d \tau\,
  {\rm tr}\big [\dot{\bm c}_{\alpha}(\tau){\bm C}(\tau)\big].
\ee
Here, ${\bm C}(t)\equiv\{C_{uv}(t)\}$ and
\be
 C_{uv}(t)\equiv\la\hat Q_{u}(t)\hat Q_v\textbf{}(0)\ra.
\ee
This is the time--domain Meir--Wingreen's formula\cite{Mei922512} for general systems.
This formula can be derived via such as the nonequilibrium Green's function method \cite{Oja08155902} or the dissipaton thermofield approach.\cite{Wan22044102}
In the frequency domain, it can be recast as
\be\label{gfr1}
J_{\alpha}
=\frac{2}{\pi}\sum_{uv}\!\int_{-\infty}^{\infty}\!\!{\rm d}\w\,\frac{\w\Phi^{}_{\alpha uv}(\w)}{1-e^{-\beta_{\alpha}\w}}
     {\rm Re}[\wti C_{vu}(-\w)],
\ee
with the hybrid bath spectral density
\be\label{Phidef}
\Phi^{}_{\alpha uv}(\w)\equiv \frac{1}{2i}\int_{-\infty}^{\infty}\!\!{\d}t\,e^{i\w t}\phi^{}_{\alpha uv}(t),
\ee
and the system correlation resolution
\be\label{spec}
\wti C_{uv}(\w)\equiv\int_{0}^{\infty}\!\!{\d}t\,e^{i\w t}C_{uv}(t).
\ee
The hybrid bath correlation function is related to its spectral density via
the fluctuation--dissipation theorem\cite{Wei21,Yan05187}
\be
	c_{\alpha uv}(t)
=\frac{1}{\pi}\!\int_{-\infty}^{\infty}\!\!{\rm d}\w\,\frac{e^{-i\w t}\Phi^{}_{\alpha uv}(\w)}{1-e^{-\beta_{\alpha}\w}}.
\ee

To evaluate \Eq{gfr1} for the present multimode BO system,
we propose a discrete imaginary--frequency (DIF) method.
This method is based on three ingredients as follows.
(\emph{i}) The exponential decomposition of the hybrid bath correlation functions (with $\gamma_{\as k}$ assumed real),
\begin{align}\label{dis}
c_{\alpha u v}(t)=\sum_k g_{\alpha ku v} e^{-\gamma_{\alpha k} t}.
\end{align}
This can be readily achieved with some sum--over--pole schemes
\cite{Hu10101106,Hu11244106,Din11164107,Din12224103,Cui19024110,Zha20064107}
or the time--domain Prony fitting decomposition scheme.\cite{Che22221102}
(\emph{ii}) The relation of entangled system--bath correlations [cf.\,\Eq{app_lang}]
\begin{align}\label{S155}
\ddot{C}_{u v}(t) &=-\Omega_{u}\sum_{u'} V_{uu'} C_{u'v}(t)-\Omega_{u}\sum_\alpha X_{\alpha u v}(t)
\nl &\quad +\Omega_{u}\sum_{\alpha u'}\iit\phi_{\alpha u u'}(t-\tau) C_{u'v}(\tau),
\end{align}
where $X_{\as uv}(t)\equiv\langle\hat{F}_{\alpha u}^{\B}(t) \hat{Q}_v(0)\rangle$.
In terms of the Laplace transform
$\wti f(\omega)\equiv\int_{0}^{\infty}{\rm d}t\,e^{i\omega t}f(t)$, it reads
\begin{align}\label{CX}
	\wti{\boldsymbol{C}}(\omega)&=\bigg[\bo\boldsymbol{V}-\omega^2 \mathbf{I}-\bo\sum_\alpha{\wbpi}_\alpha(\omega)\bigg]^{-1}\bigg[\dot{\bC}(0)-i\omega\bC(0)
\nl &\quad -\bo\sum_\alpha \wti{\boldsymbol{X}}_\alpha(\omega)\bigg].
\end{align}
(\emph{iii})
The system--bath entanglement theorem [cf.\,the Eq.(17) of Ref.\,\onlinecite{Wan22044102}]
\begin{align}\label{X1}
\boldsymbol{X}_{\alpha}(t)=2\operatorname{Im} \int_0^{\infty}\!\!\!\!\mathrm{~d}\tau\,
     {\bm c}^T_{\alpha}(t+\tau){\bm C}^T(\tau).
\end{align}
Here, (\emph{ii}) is from the BO algebra while (\emph{iii}) exists for general systems.

Substituting \eq{dis} into \eq{X1}, we obtain the following frequency--domain expression
\begin{align}\label{X}
	\wti{\bm{X}}_{\as}(\omega)=\sum_{k}\frac{2}{\gamma_{\as k}-i\omega}
       \operatorname{Im}\big[\wti{\bC}(i\gamma_{\as k})\bm{g}_{\as k}\big]^{T}.
\end{align}
We find that once $\{\wti{\bC}(i\gamma_{\as k})\}$ is known,
$\wti{\bm{X}}_{\as}(\omega)$ [\Eq{X}] and hence $\wti{\boldsymbol{C}}(\omega)$ [\Eq{CX}] will be readily obtained.
On the other hand, $\{\wti{\bC}(i\gamma_{\as k})\}$ can be solved by \eqs{CX} and (\ref{X}) in a self-consistent manner, i.e.
\begin{subequations}
	\begin{align}
	&\wti{\boldsymbol{C}}(i\gamma_{\as' k'})=\bigg[\bo\boldsymbol{V}+\gamma^2_{\as' k'}\mathbf{I}-\bo\sum_\alpha{\wbpi}_\alpha(i\gamma_{\as' k'})\bigg]^{-1}\bigg[\dot{\bC}(0)
    \nl &\qquad\qquad\qquad +\gamma_{\as' k'}\bC(0)-\bo\sum_\alpha \wti{\boldsymbol{X}}_\alpha(i\gamma_{\as' k'})\bigg],\\
	&\wti{\bm{X}}_{\as}(i\gamma_{\as' k'})=\sum_{k}\frac{2}{\gamma_{\as k}+\gamma_{\as' k'}}
       \operatorname{Im}\big[\wti{\bC}(i\gamma_{\as k})\bm{g}_{\as k}\big]^{T}.
	\end{align}
\end{subequations}
Note that for the multimode BO system in this paper, $\phi_{\alpha uv}=\phi_{\alpha vu}$.
The steady--state heat current can then be obtained from \Eq{gfr1} via
\begin{align}\label{gfr2}
J_{\alpha}&
=-\frac{2}{\pi}\sum_{uv}\!\int_{-\infty}^{\infty}\!\!{\rm d}\w\,\frac{\w\Phi^{}_{\alpha uv}(\w)}{e^{\beta_{\alpha}\w}-1}
     {\rm Re}[\wti C_{vu}(\w)].
\end{align}

By far, we have proposed two different methods to evaluate the steady--state heat current
of the multimode BO system. One is the algebraic method with the aid of the EOM of the density matrix.
The other is the DIF method followed by the Meir--Wingreen's formula.
These two exact methods are numerically consistent in our calculations.
As tested, the simulation results are also found consistent with  those via the dissipaton theory. \cite{Wan20041102, Wan22044102}
Note that there is $\sum_\alpha J_{\alpha}=0$ in the steady state,
which is also confirmed in our calculations.

\begin{figure}[h]
\includegraphics[width=0.5\textwidth]{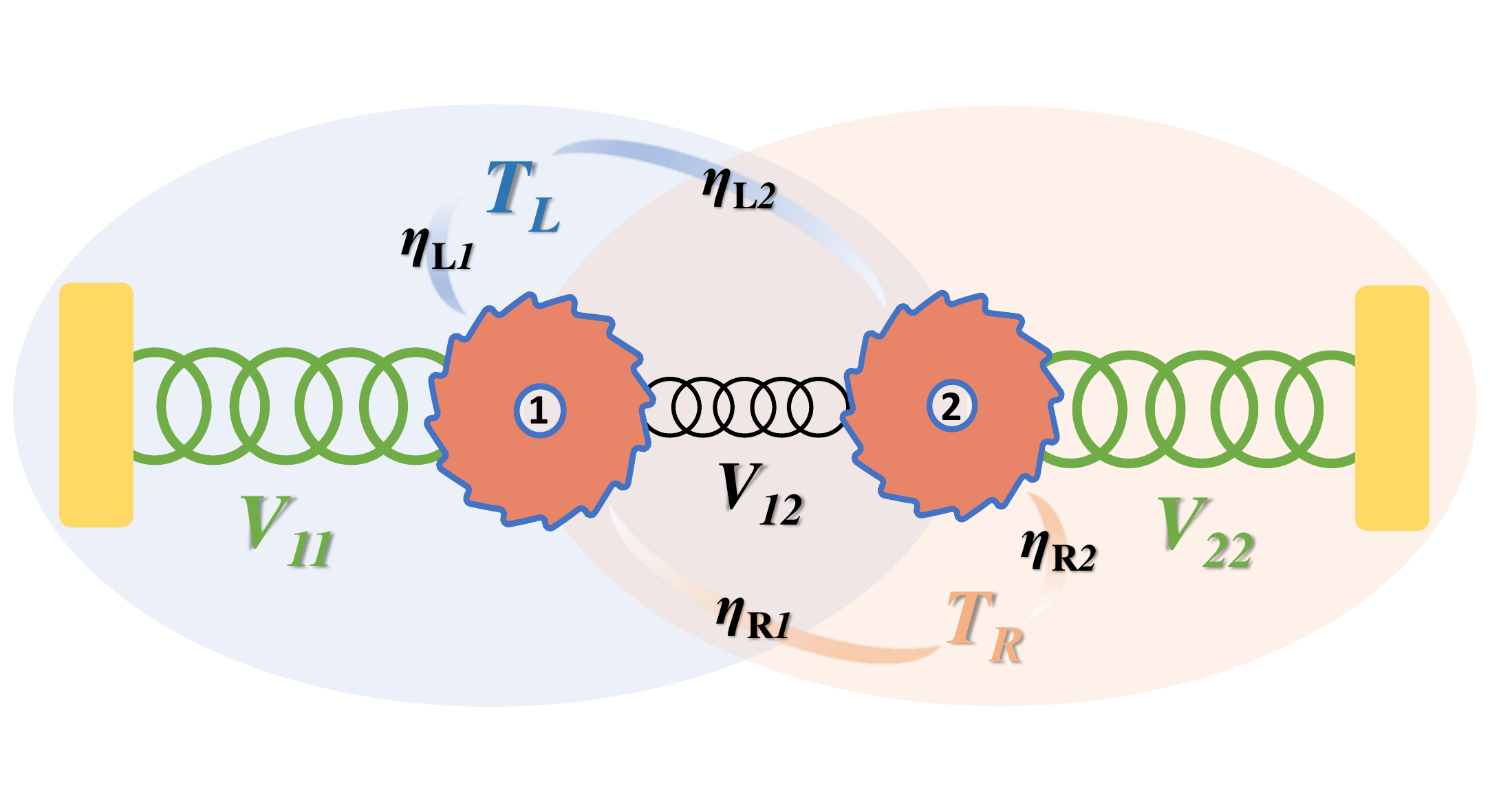}
\caption{Schematic diagram of the model for numerical demonstrations.}\label{fig1}
\end{figure}

\section{Numerical demonstrations}
\label{num}
For numerical demonstrations, we adopt [cf.\,\eq{Phidef}]
\begin{align}\label{BO}
{\bm{\Phi}}_{\alpha}(\w)={\bm{\eta}}_{\as}{\rm Im}\frac{\Omega_{\B}^{2}}{\Omega_{\B}^{2}-\omega^{2}-i\omega\gamma_{\B}},
\end{align}
with $ \bm{\eta}_{\as}\equiv\{\eta_{\as uv};\eta_{\as uv}=\eta_{\as u}\delta_{uv}\} $ specifying the system--bath coupling strengths.
Exemplified are the calculations on transient dynamics and heat currents.
Selected is a two--mode BO system embedded in two (left and right with $\alpha$=L and R respectively)
reservoirs at different temperatures.
The schematic diagram of the model is shown in \Fig{fig1}.
From an application point of view, this setup can be experimentally realized via, for example, molecular junctions.\cite{Seg05034301,Gia06217,Sai07027203,Bel081457}
The system--bath coupling strength $\{\eta_{\alpha u}\}$ can be experimentally adjusted via manipulating the distances between molecules.
We select the parameters $\Omega_{1}=\Omega_{2}=V_{11}=V_{22}=\Omega_{\B}=1600$ cm$^{-1}$
and $\gamma_{\B}=4\Omega_{\B}$.
Other parameters are specified in the caption of each figure.

\begin{figure}[h]
\includegraphics[width=0.48\textwidth]{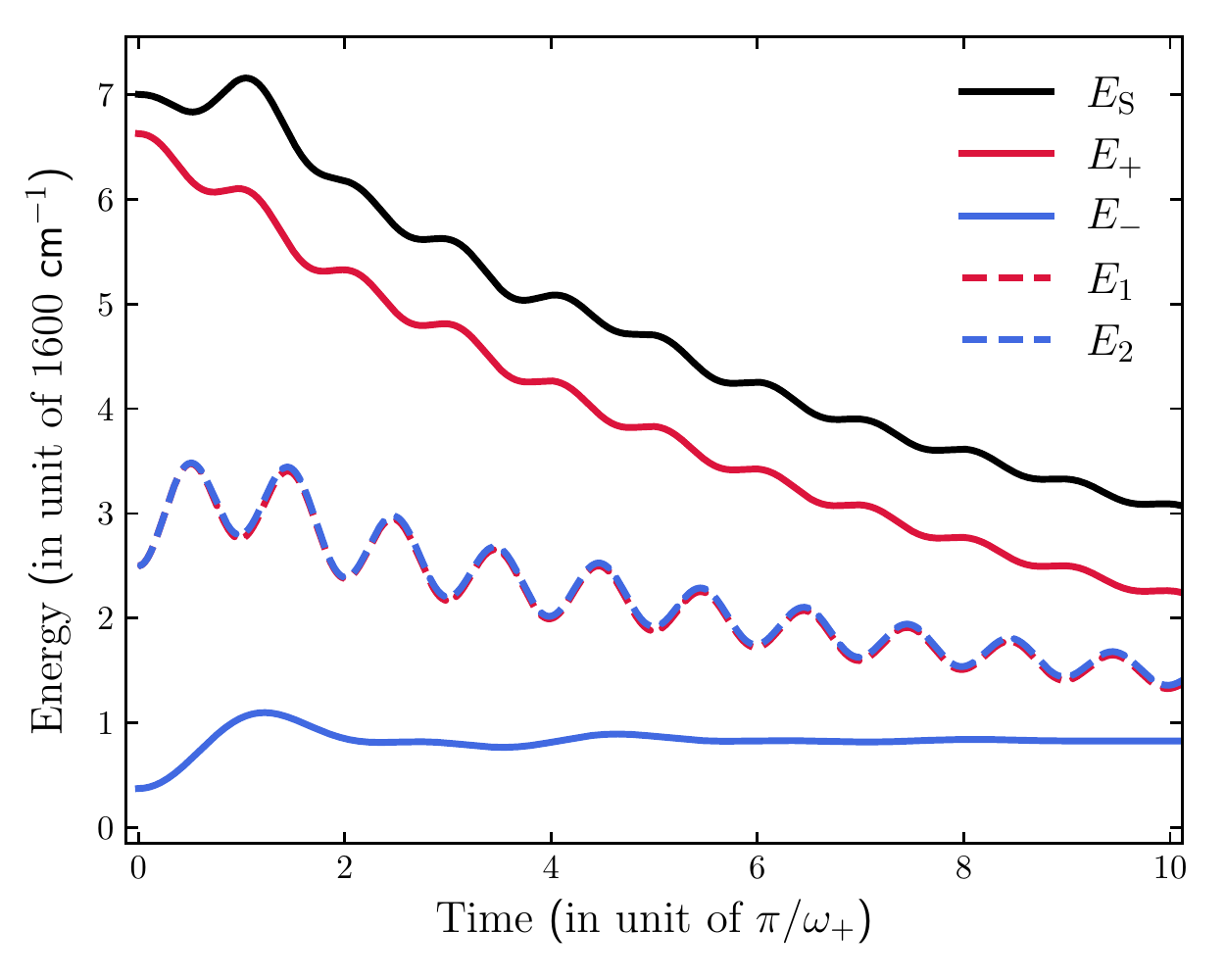}
\caption{Evaluations on the transient two--mode BO dynamics in terms of the energies of the system oscillators. See text for details.
Parameters are selected as $V_{12}=V_{21}=800$cm$^{-1}$, $\eta_{{\rm L}1}=\eta_{{\rm R}2}=320$cm$^{-1}$, $\eta_{{\rm L}2}=\eta_{{\rm R}1}=160$cm$^{-1}$, $T_{\rm L}=50$K and $T_{\rm R}=500$K.}\label{fig2}
\end{figure}

\begin{figure}[h]
\includegraphics[width=0.48\textwidth]{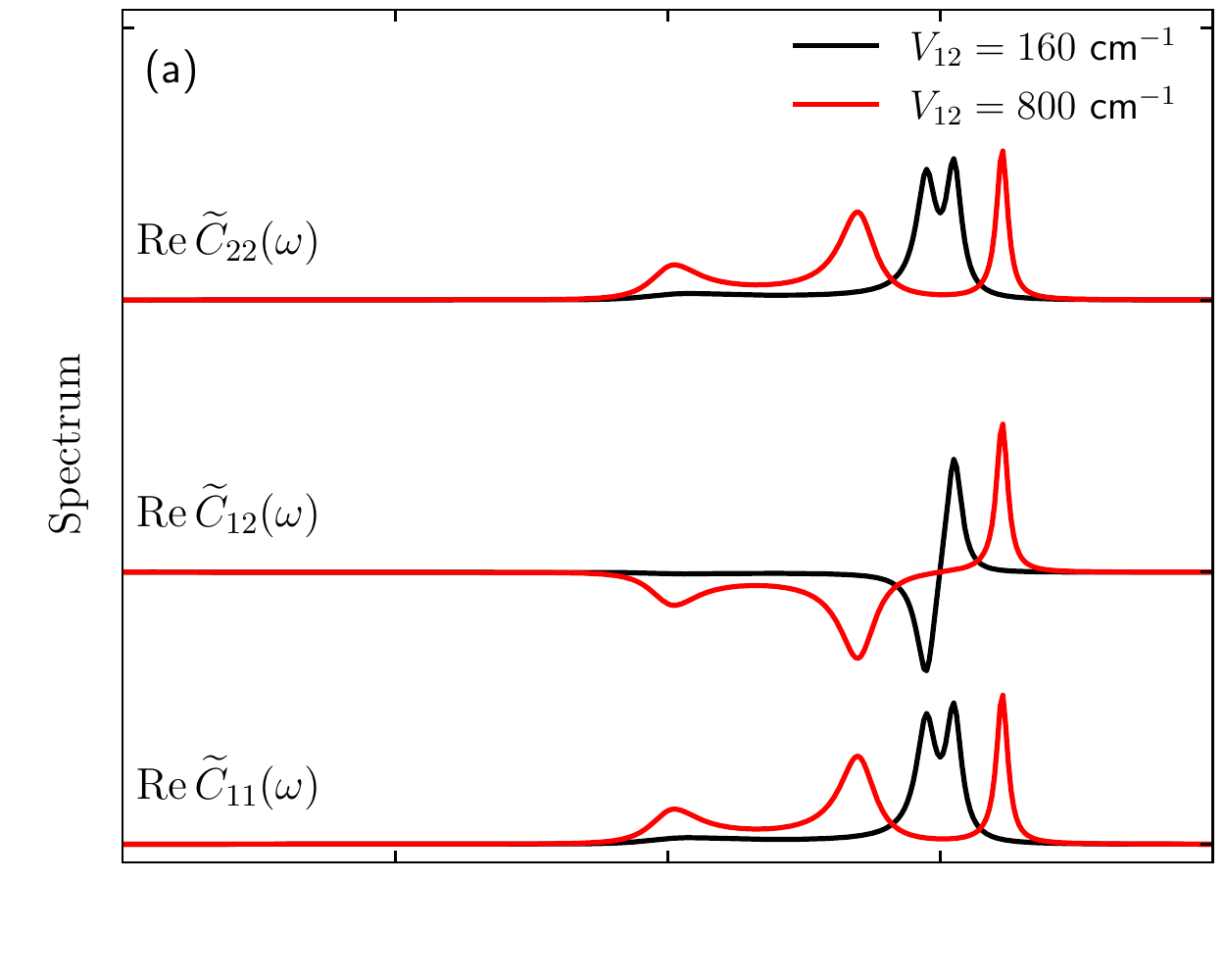}
 \vspace{-2em}
\includegraphics[width=0.48\textwidth]{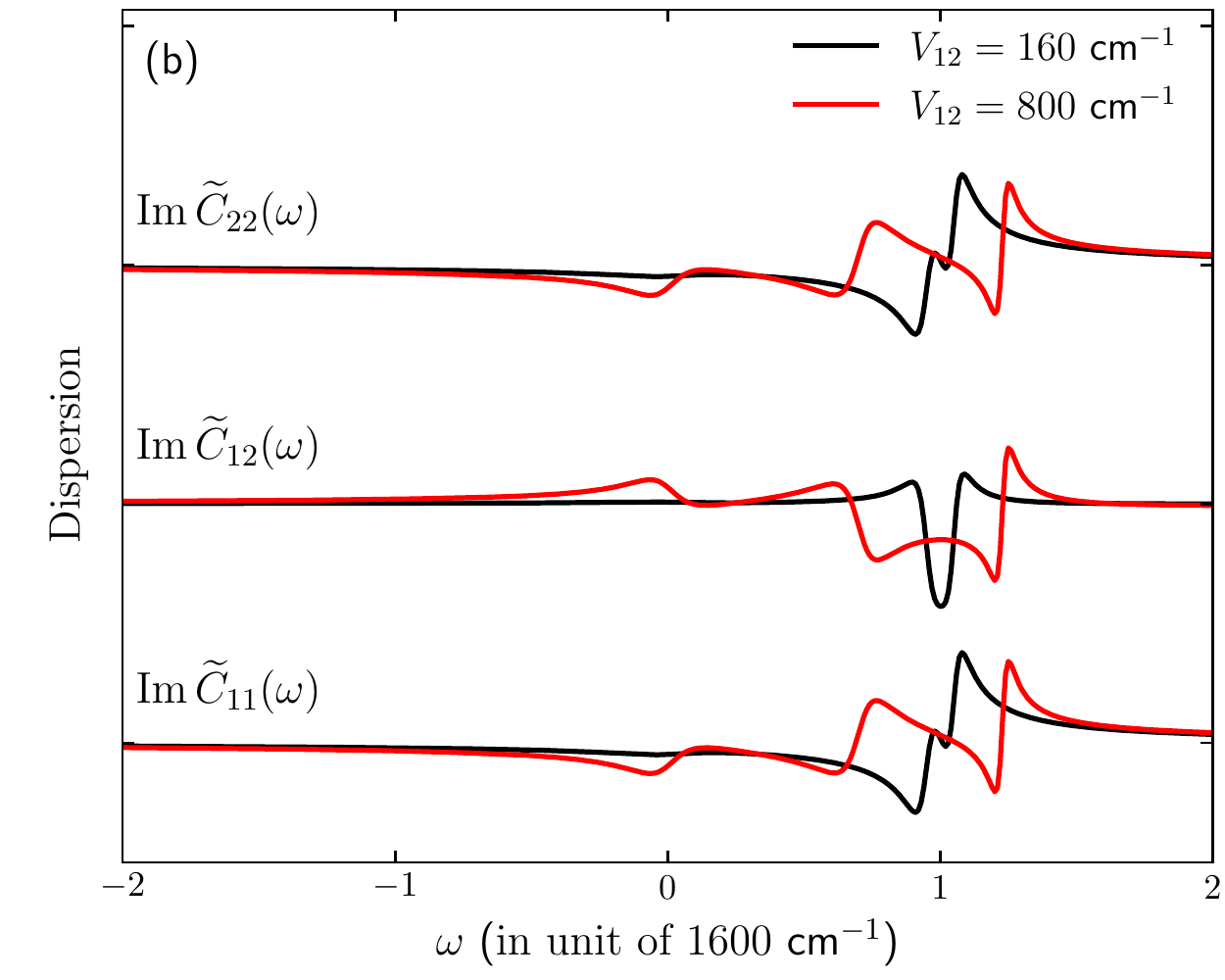}
\caption{The $\wti{\boldsymbol{C}}(\omega)$ evaluated with $V_{12}=V_{21}=160$cm$^{-1}$ (black) and $800$cm$^{-1}$ (red) at $T_{\rm L}=T_{\rm R}=275$K.
Other parameters are the same as in \Fig{fig2}.}\label{fig3}
\end{figure}

\begin{figure}[h]
\includegraphics[width=0.48\textwidth]{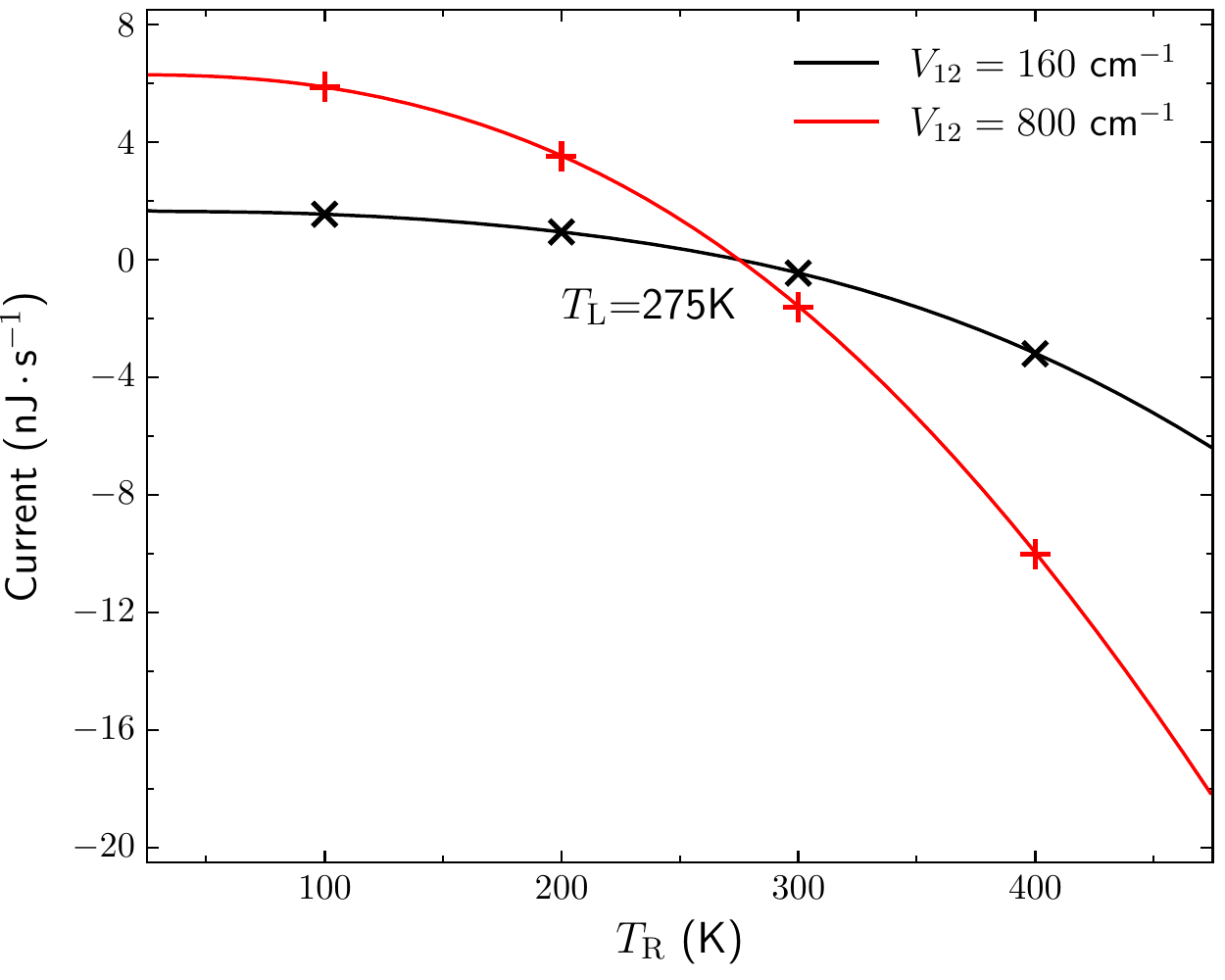}
\caption{Heat current versus temperature, with $V_{12}=V_{21}=160$cm$^{-1}$ (black) and $800$cm$^{-1}$ (red).
The temperature of the right bath, $T_{\rm R}$, varies from 25K to 475K
while that of the left bath, $T_{\rm L}$, is fixed at 275K. Other parameters are the same as in \Fig{fig2}.
The crosses marked along the curves are results from the DIF method.
}\label{fig4}
\end{figure}

Figure \ref{fig2} depicts the transient dynamics of the system energies,
$
 E_{\tS}=\la H_{\tS} \ra
$,
$E_{i=1,2}=(\Omega_i/2) \la\hat P_i^2\ra+(V_{ii}/2)\la\hat Q_i^2\ra$,
and $E_{\pm}=\la\hat p_{\pm}^2\ra/2+(\omega_{\pm}^2/2)\la\hat q_{\pm}^2\ra$
of the normal modes.  
We select $V_{12}=V_{21}=800$cm$^{-1}$ resulting in $\omega_{+}=1960$cm$^{-1}$ and $\omega_{-}=1130$cm$^{-1}$.
The $\omega_{+}$ plays the dominate role in realistic observations.
The system is initially at the factorized state, $\rho_{\tS}(0)\otimes\rho^{\rm cano}_{\B}$,
with $\rho_{\tS}(0)=\otimes_{i=1,2}\rho_{i}$
where $\rho_{i}$ is the ground state of the $i$th isolated oscillator
but with a dimensionless position displacement of $d=-2$.
To be specific, $\rho_{i}=|\psi^{(i)}\ra\la \psi^{(i)}|$, with
\[
\psi^{(i)}(Q_i)=\bigg(\frac{\sqrt{V_{ii}/\Omega_i}}{\pi}\bigg)^{1/4}\exp\bigg[-\frac{\sqrt{V_{ii}/\Omega_i}}{2}(Q_i+2)^2\bigg].
\]
Apparently, $E_{\tS}=E_{+}+E_{-}\neq E_{1}+E_{2}$.

The two panels of \Fig{fig3} show the spectra and dispersions, i.e.\
the real and imaginary parts of $\wti{\boldsymbol{C}}(\omega)$, respectively,
evaluated via the DIF method proposed in \Sec{thsec3b}.
Obviously,
the splitting of peaks becomes larger when $V_{12}=V_{21}=800$cm$^{-1}$,
compared to the case of $V_{12}=V_{21}=160$cm$^{-1}$.
With the equal temperature of both reservoirs, the detailed--balance relation has been satisfied in our evaluations.

Shown in \Fig{fig4} is the steady--state heat current with
the temperature of the right bath, $T_{\rm R}$, varying from 25K to 475K
while that of the left bath, $T_{\rm L}$, fixed at 275K.
Apparently, the current increases with the inter-mode coupling strength, $V_{12}$.
Also the current changes direction at the condition $T_{\rm R}=T_{\rm L}$.
The differential conductance is found of similar behavior which is thus not shown explicitly.
From \Fig{fig4}, one can observe that there is an asymmetry in magnitude of current
between two scenarios of $T_R < T_L$ and $T_R > T_L$ with the fixed temperature difference.
The mechanism can be interpreted as follows.
The heat current increases with  not only the temperature difference but also the heat capacity.
The phonon frequency involved in our simulation is around $1600$cm$^{-1}$.
The corresponding Einstein temperature is over $2000$K.
For the temperature range in \Fig{fig4}, the heat capacity increases with the temperature, leading to the observed asymmetry.%

\section{Summary}\label{thsec4}
   To conclude, we develop an algebraic method to study multimode Brownian oscillators connected to multiple reservoirs at different temperatures.
The algebraic approach directly gives the exact time--local EOM for the reduced density operator.
Based on this approach,
 we can easily extract not only the reduced system but also the hybrid bath dynamical informations.
We exploit the EOM to study the heat transport problem, i.e.\
the heat current from the reservoir
to the local impurity system.
%
On the other hand, we also adopt another different approach, the Meir--Wingreen's formula which is for general systems,
together with the system--bath entanglement theorem to evaluate the steady--state heat current.
For its application to the BO system,
we propose a discrete imaginary--frequency (DIF) self-consistent evaluation scheme.
Both approaches can serve as initial steps to systematically include anharmonic effects.
For more generic (non-integrable) bosonic setups,
the self-energy term needs to be modified, \cite{Du20034102} which can be obtained
via self-consistent iteration approximately.
Work in this direction is in progress.
It is anticipated that the methods presented in this work would
constitute basic components for nonequilibrium statistical mechanics of open quantum systems.
%
Moreover,
it is still difficult to  compute the fluctuation of heat current under the present theoretical framework of Eqs.(\ref{e67})--(\ref{defop1}).
However, it may be achieved
with the help of dissipaton theory.\cite{Wan20041102, Wan22044102}
This constitutes another direction of further development.

\begin{acknowledgements}
	Support from the Ministry of Science and Technology of China (Grant No.\ 2021YFA1200103) and the National Natural Science Foundation of China (Grant Nos.\ 22103073 and 22173088) is gratefully acknowledged.
\end{acknowledgements}

\appendix*
\section{Related BO algebra and BO properties}
The EOM (\ref{e67})--(\ref{defop1}) can be established by the Yan--Mukamel method.\cite{Yan885160,Xu037,Xu09074107}
Here we present another approach via undetermined coefficients.
Both methods are on the basis of the EOM of the first and second order moments,
which fully characterize the Gaussian wave packet (GWP).
Let us start from the Heisenberg EOM for any operator, $\dot{\hat{O}}(t)=i[H_{\T},\hat O(t)]$.
We have
\bsube\label{Appa1}
\begin{align}\label{Appa1a}
 \dot{\hat Q}_u(t)&=\Omega_u{\hat P}_u(t),\\
 \dot{\hat P}_u(t)&=-\sum_vV_{uv}{\hat Q}_v(t)-\sum_{\alpha}\hat F_{\alpha u}(t).\label{Appa1b}
\end{align}
\esube
The solution to 
$ \hat F_{\alpha u}(t)\equiv e^{iH_{\T}t}\hat F_{\alpha u}e^{-iH_{\T}t} $
can be obtained as
[cf.\,the Eq.\,(7) of Ref.\,\onlinecite{Du20034102}]
\begin{align}\label{FB}
\hat F_{\alpha u}(t)=\hat F^{\B}_{\alpha u}(t)-\sum_{v}\iit \phi_{\alpha u v}(t-\tau)\hat Q_{v}(\tau).
\end{align}
From \Eq{Appa1} we obtain
\bsube\label{Appa2}
	\begin{align}
	\dot{\chi}^{QQ}_{uv}(t)&=\Omega_u\chi^{PQ}_{uv}(t),\\
	\dot{\chi}^{QP}_{uv}(t)&=\Omega_u\chi^{PP}_{uv}(t),
	\end{align}
\esube
and
\bsube\label{Appa3}
	\begin{align}
	\dot{\chi}^{PQ}_{uv}(t)&=-\sum_{v'}V_{uv'}\chi^{QQ}_{v'v}(t)-i\sum_{\alpha}\langle [\hat{F}_{\alpha u}(t),\hat{Q}_{v}(0)]\rangle, \\
    \dot{\chi}^{PP}_{uv}(t)&=-\sum_{v'}V_{uv'}\chi^{QP}_{v'v}(t)-i\sum_{\alpha}\langle [\hat{F}_{\alpha u}(t),\hat{P}_{v}(0)]\rangle.
	\end{align}
\esube
Substituting \Eq{FB} into \Eq{Appa3}, we get
\bsube\label{Appa4}
	\begin{align}
	\dot{\chi}^{PQ}_{uv}(t)&=-\sum_{v'}V_{uv'}\chi^{QQ}_{v'v}(t)\nl &\qquad +\sum_{\alpha v'}\iit\phi_{\alpha uv'}(t-\tau)\chi^{QQ}_{v'v}(\tau),\\
	\dot{\chi}^{PP}_{uv}(t)&=-\sum_{v'}V_{uv'}\chi^{QP}_{v'v}(t)\nl &\qquad +\sum_{\alpha v'}\iit\phi_{\alpha uv'}(t-\tau)\chi^{QP}_{v'v}(\tau).
	\end{align}
\esube

The traditional Langevin equation can be recovered by substituting \Eq{FB} into \Eq{Appa1} as
\begin{align}\label{app_lang}
\ddot{\hat Q}_u(t)&=-\Omega_u\sum_vV_{uv}{\hat Q}_v(t)-\Omega_u\sum_{\alpha}\hat F^{\B}_{\alpha u}(t)
\nl &\quad+\Omega_u\sum_{\alpha v}\iit \phi_{\alpha u v}(t-\tau)\hat Q_{v}(\tau).
\end{align}
The solutions to it, i.e.\ \Eq{nonlocal} with \Eq{tmatirx},
can be proved by \eqs{Appa2} and (\ref{Appa4}) together with initial values
$\bm{\chi}(0)=\ddot{\bm{\chi}}(0)=\bf{0}$ and $\dot{\bm{\chi}}(0)=\bm{\Omega}$.
Furthermore, \eqs{Appa2} and (\ref{Appa4}) can be resolved in terms of the Laplace transform
$\wti f(\omega)\equiv\int_{0}^{\infty}{\rm d}t\,e^{i\omega t}f(t)$ as
\bsube
\begin{align}
  \wti\cc_{QQ}(\omega)&=[\bo\bv-\omega^2{\bf I}-\bo\sum_\alpha{\wbpi}_{\alpha}(\omega)]^{-1}\bo,\\
  \wti\cc_{QP}(\omega)&=i\omega\wti\cc_{QQ}(\omega)\bo^{-1},\\
  \wti\cc_{PQ}(\omega)&=-i\omega\bo^{-1}\wti\cc_{QQ}(\omega),\\
  \wti\cc_{PP}(\omega)&={\bo}^{-1}+\omega^2{\bo}^{-1}\wti\cc_{QQ}(\omega){\bo}^{-1}.
\end{align}
\esube
Equations (\ref{Appa2}) and (\ref{Appa4}) can be recast into the matrix form as
\begin{align}\label{Appa6}
\dot{\bt}(t)=
\begin{bmatrix}
\bf{0} & \bo  \\   -\bv        & \bf{0}
\end{bmatrix}\bt(t)+\sum_{\alpha}       \iit
\begin{bmatrix}
\bf{0} & \bf{0}       \\   \bp(t-\tau) & \bf{0}
\end{bmatrix}\bt(\tau).
\end{align}
By the first identity of \Eq{lammda}, we can recast the $ \bm{\Lambda}(t) $ matrix using \Eq{Appa6} as
\begin{align}\label{Appa7}
{\bm\Lambda}(t)&=\begin{bmatrix}
\bf{0} & \bo  \\   -\bv        & \bf{0}
\end{bmatrix}+\sum_{\alpha}             \iit
\begin{bmatrix}
\bf{0} & \bf{0}       \\   \bp(t-\tau) & \bf{0}
\end{bmatrix}\bt(\tau)\bt^{-1}(t).
\end{align}
Equation (\ref{Gamma_contri}) is resulted from the comparison of \Eq{Appa7} to the second identity of \Eq{lammda}
with \Eq{separate}.


Consider the initial factorized state before the system and baths interact, i.e.
\begin{align}\label{app_ini}
\rho_{\T}(0)=\rho_{\s}(0)\otimes\prod_{\alpha}\rho^{\B}_{\alpha}(\beta_{\as}).
\end{align}
We have denoted the average $ \langle  (\,\cdot\,) \rangle_{\ini} $ over
the total composite state with the above initial total GWP.
That is in the Schr\"{o}dinger picture, $\langle\hat O\rangle_{\ini}={\rm Tr}[\hat O\rho_{\T}(t)]$,
which is equal to $\langle\hat O\rangle_{\ini}={\rm Tr}[\hat O(t)\rho_{\T}(0)] $ in the Heisenberg picture.
From \eq{local}, the EOM of the first order moments,
\be \label{1st_def}
   Q_{u}(t)\equiv \big\langle \hat{Q}_{u}(t)  \big\rangle_{\ini}\ \ \text{and}\ \ P_{u}(t)\equiv \big\langle \hat{P}_{u}(t)  \big\rangle_{\ini},
\ee
can be obtained as
\begin{align}\label{B2}
\begin{bmatrix} \dot{\bm{ Q}}(t) \\ \dot{\bm{P}}(t) \end{bmatrix}= {\bm\Lambda}(t)
\begin{bmatrix} \bm{ Q}(t) \\ \bm{P}(t) \end{bmatrix}.
\end{align}
Note that $ \big\langle \bm{\hat{F}}^{\eff}_{\alpha}(t)  \big\rangle_{\ini}= \big\langle \bm{\hat{F}}^{\B}_{\alpha}(t)  \big\rangle_{\ini}=\bm{0} $.  
Actually for any system operator $\hat A$, there is
$ \big\langle \bm{\hat{F}}^{\eff}_{\alpha}(t)\hat A(0)  \big\rangle_{\ini}
= \big\langle \bm{\hat{F}}^{\B}_{\alpha}(t)  \hat A(0)  \big\rangle_{\ini}=\bm{0} $.

The EOM of the second order moments,
\begin{subequations}\label{B3}
	\begin{align}
    &W^{QQ}_{uv}(t)\equiv\frac{1}{2}\big\langle \{\delta \hat{Q}_{u}(t) ,\delta \hat{Q}_{v}(t)  \}\big\rangle_{\ini}\,,\\
	&W^{QP}_{uv}(t)\equiv\frac{1}{2}\big\langle \{\delta \hat{Q}_{u}(t) ,\delta \hat{P}_{v}(t) \} \big\rangle_{\ini}=W^{PQ}_{vu}(t)\,,\\
	&W^{PP}_{uv}(t)\equiv\frac{1}{2}\big\langle\{ \delta \hat{P}_{u}(t) ,\delta \hat{P}_{v}(t)  \}\big\rangle_{\ini}\,,
	\end{align}
\end{subequations}
with $ \delta \hat{Q}_{u}(t) \equiv \hat{Q}_{u}(t)-Q_{u}(t)  $ and $ \delta \hat{P}_{u}(t) \equiv \hat{P}_{u}(t)-P_{u}(t)  $
can also be derived from \eq{local}.
For convenience, we recast \eq{B3} in the matrix form as
\begin{align}\label{W}
\bW(t)&\equiv
\begin{bmatrix}
\bm{W}^{QQ}(t) & \bm{W}^{QP}(t)\\
\bm{W}^{PQ}(t) & \bm{W}^{PP}(t)
\end{bmatrix}
\nl &=\frac{1}{2}\bigg\langle  \bigg\{
\begin{bmatrix}
\delta  \bm{\h Q}(t)\\
\delta  \bm{\h P}(t)
\end{bmatrix},
\begin{bmatrix}
\delta  \bm{\h Q}(t)\\
\delta  \bm{\h P}(t)
\end{bmatrix}^{T}
\bigg\} \bigg\rangle.
\end{align}
We obtain from \eq{local} that 
\begin{align}\label{appbdotW}
\dot{\bW}(t)
&=\bm{\Lambda}(t)\bW(t)
+\bW(t)\bm{\Lambda}^{T}(t)
\nl &\qquad +\sum_\alpha\begin{bmatrix}
\ \bf{0}\ &\ \big[\bm{\zeta}_{\asb}(t)\big]^{T}\ \\
\ \bm{\zeta}_{\asb}(t)\ &\ {\bm{\tilde\zeta}}_{\asb}(t)+\big[{\bm{\tilde\zeta}}_{\asb}(t)\big]^{T}\
\end{bmatrix},
\end{align}
where
\begin{subequations}\label{B6}
	\begin{align}
	&\bm{\zeta}_{\asb}(t)        =-\operatorname{Re}\big\langle\bm{\hat{F}}^{\rm eff}_{\alpha}(t)\bm{\h Q}^{T}(t)\big\rangle_{\ini} \,,\\
	&{\bm{\tilde\zeta}}_{\asb}(t)=-\operatorname{Re}\big\langle\bm{\hat{F}}^{\rm eff}_{\alpha}(t)\bm{\h P}^{T}(t)\big\rangle_{\ini} \,.
	\end{align}
\end{subequations}

The evaluations of $\bm{\zeta}_{\asb}(t)$ and ${\bm{\tilde\zeta}}_{\asb}(t)$
 can be done by substituting \eqs{nonlocal} and (\ref{Feff_vc}) into \Eq{B6}.
Denoting
\begin{align}\label{appbzeta}
\bm{\xi}_{\asb}(\tau;t)&\equiv\bm{r}_{\asb}(\tau)
     +\bm{\wti V}(t)\iitp \bm{\chi}(\tau') \bm{r}_{\asb}(\tau-\tau')
\nl &\qquad     +\bm{\Gamma}(t)\iitp \bo^{-1}\dot{\cc}(\tau')\bm{r}_{\asb}(\tau-\tau')
\nl &\qquad     +\bo^{-1}\iitp \ddot{\cc}(\tau')\bm{r}_{\asb}(\tau-\tau'),
\end{align}
we have
\begin{align}\label{zetaapp}
 \begin{bmatrix}  \bm{\zeta}_{\asb}(t) &  {\bm{\tilde\zeta}}_{\asb}(t)   \end{bmatrix}
&=\iit    \bm{\xi}_{\asb}(\tau;t) \begin{bmatrix}  \cc^T(\tau) &  \dot\cc^T(\tau)\bo^{-1}   \end{bmatrix}.
\end{align}
Equation (\ref{zetafinal}) can now be obtained by noticing the following relation of double integrals
\begin{align}
  &\iit\iitp f(\tau')g(\tau-\tau')h(\tau)\nl =&\iit
     \int_{0}^{\tau}\!{\rm d}\tau'\, [f(\tau')g(\tau-\tau')h(\tau)
       \nl &\qquad\qquad\qquad +f(\tau)g(\tau'-\tau)h(\tau')].
\end{align}
Note in \eqs{zetafinal}--(\ref{bu})
\begin{align}
	\dot\bu_{\alpha}(t)&=\int_{0}^{t}\!{\rm d}\tau\,\dot\cc(\tau) \bm{r}_{\asb}(t-\tau),\\
	\ddot\bu_{\alpha}(t)&=\bm{\Omega}\bm{r}_{\asb}(t)+\int_{0}^{t}\!{\rm d}\tau\,\ddot\cc(\tau) \bm{r}_{\asb}(t-\tau).
\end{align}

Turn now to the EOM of the reduced density operator ${\rho}_{\s}(t)$.
Concerning the Gaussian property and the Liouville-von Neumann equation for the total density operator,
$ \dot{\rho}_{\T}(t)=-i[H_{\T},\rho_{\T}(t)] $, the desired EOM is of the form
\begin{align}\label{appbmotionrho}
\dot{\rho}_{\s}(t)&=-i\big[H_{\s},\rho_{\s}(t)\big]
+\sum_{\alpha uv}\hat{Q}_{u}^{\obarminus}
\big[a^{\obarplus}_{\alpha uv}(t)\hat{Q}_{v}^{\obarplus} +b^{\obarplus}_{\alpha uv}(t)\hat{P}_{v}^{\obarplus}
\nl &\qquad
    +a^{\obarminus}_{\alpha uv}(t)\hat{Q}_{v}^{\obarminus}+b^{\obarminus}_{\alpha uv}(t)\hat{P}_{v}^{\obarminus}\big]\rho_{\s}(t),
\end{align}
with the coefficients $\{a^{\obarplus}_{\alpha uv}\}$, $\{b^{\obarplus}_{\alpha uv}\}$, $\{a^{\obarminus}_{\alpha uv}\}$, and $\{b^{\obarminus}_{\alpha uv}\}$ to be determined.
To do that, consider an arbitrary system operator $\hat A$.
We can obtain the EOM of $\langle\hat A\rangle_{\ini}={\rm Tr}_{\s}[\hat A\rho_{\s}(t)]$ from \Eq{appbmotionrho} as
\begin{align}\label{appbmotionA}
\frac{\rm d}{{\rm d}t}\langle\hat A\rangle_{\ini}&=i\big\langle\big[H_{\s},\hat A\big]\big\rangle_{\ini}
-\sum_{\alpha uv}\big\langle
\big[a^{\obarplus}_{\alpha uv}(t)\hat{Q}_{v}^{\obarplus} +b^{\obarplus}_{\alpha uv}(t)\hat{P}_{v}^{\obarplus}
\nl &\qquad
    -a^{\obarminus}_{\alpha uv}(t)\hat{Q}_{v}^{\obarminus}-b^{\obarminus}_{\alpha uv}(t)\hat{P}_{v}^{\obarminus}\big]\hat{Q}_{u}^{\obarminus}\hat A\big]\big\rangle_{\ini}.
\end{align}
We can now derive the EOM of the first and second order moments from \Eq{appbmotionA} and compare
the results to \eqs{B2} and (\ref{appbdotW}). After some simple steps,
the coefficients $\{a^{\obarplus}_{\alpha uv}\}$, $\{b^{\obarplus}_{\alpha uv}\}$, $\{a^{\obarminus}_{\alpha uv}\}$, and $\{b^{\obarminus}_{\alpha uv}\}$ can be determined
as $\{\wti\Gamma_{\alpha uv}(t)\}$, $\{\Gamma_{\alpha uv}(t)\}$,
$\{\wti\zeta_{\alpha uv}(t)\}$, and $-\{\zeta_{\alpha uv}(t)\}$, respectively.
Thus the final EOM of ${\rho}_{\s}(t)$ is obtained as \eqs{e67}--(\ref{defop1}).
The effect of the initial state [\Eq{app_ini}] vanishes when $t\rightarrow\infty$.
The asymptotic behaviors of the coefficients in \Eq{varrhonew1}
can refer to Ref.\,\onlinecite{Xu09074107}.


\end{document}